\documentclass[9pt,twocolumn,twoside]{pnas-new}

\templatetype{pnasresearcharticle} 

\title{Circular swimming motility and disordered hyperuniform
state in an algae system}

\author[a]{Mingji Huang}
\author[b]{Wensi Hu}
\author[a]{Siyuan Yang}
\author[b,c]{Quan-Xing Liu}
\author[a,d,1]{H. P. Zhang}

\affil[a]{School of Physics and Astronomy and Institute of Natural Sciences,
Shanghai Jiao Tong University, Shanghai 200240, China}
\affil[b]{State Key Laboratory of Estuarine and Coastal Research, School of
Ecological and Environmental Sciences, East China Normal University,
Shanghai 200241, China}
\affil[c]{Shanghai Key Lab for Urban Ecological Processes and Eco-Restoration
\& Center for Global Change and Ecological Forecasting, School of
Ecological and Environmental Sciences, East China Normal University,
Shanghai 200241, China}
\affil[d]{Collaborative Innovation Center of Advanced Microstructures, Nanjing 210093,
China}

\setboolean{displaywatermark}{false}

\leadauthor{Huang} 

\significancestatement{Disordered hyperuniform materials suppress large scale density fluctuations like crystals and remain locally isotropic as liquids. These materials possess unique properties, such as isotropic photonic gap, and have attracted the attention of scientists from many disciplines. Here we show that circularly swimming marine algae can robustly self-organize into disordered hyperuniform states through long-range hydrodynamic interactions at air-liquid interfaces. Important properties measured from hyperuniform states can be quantitatively reproduced by a numerical model whose main parameters are obtained experimentally. Our work clearly demonstrates the possibility to create disordered hyperuniform states via hydrodynamic interactions and highlights the importance of such interactions in active matter systems.}

\authorcontributions{M.H.and H.P.Z. designed research; M.H., W.H. and S.Y. performed research; M.H., W.H. and H.P.Z. analyzed data; and M.H., W.H., S.Y., Q.X.L. and H.P.Z. wrote the paper.}
\authordeclaration{The authors declare no conflict of interest.}
\correspondingauthor{\textsuperscript{1}To whom correspondence should be addressed. E-mail: hepeng\_zhang@sjtu.edu.cn}


\begin{abstract}
Active matter comprises individually driven units that convert locally stored
energy into mechanical motion. Interactions between driven units
 lead to a variety of non-equilibrium collective phenomena in active matter. One of such phenomena is anomalously large density
fluctuations, which have been observed in both experiments and theories.
Here we show that, on the contrary, density fluctuations in active matter can also be
greatly suppressed. Our experiments are carried out with marine algae
($\it{Effrenium\ voratum}$) which swim in circles at the air-liquid
interfaces with two different eukaryotic flagella. Cell swimming generates fluid flow which leads to effective repulsions between cells in the far field. Long-range nature of
such repulsive interactions suppresses density fluctuations and generates disordered
hyperuniform states under a wide range of density conditions. Emergence of
hyperuniformity and associated scaling exponent are quantitatively
reproduced in a numerical model whose main ingredients are effective hydrodynamic
interactions and uncorrelated random cell motion. Our results demonstrate
a new form of collective state in active matter and suggest the possibility
to use hydrodynamic flow for self-assembly in active matter.
\end{abstract}

\dates{This manuscript was compiled on \today}
\doi{\url{www.pnas.org/cgi/doi/10.1073/pnas.XXXXXXXXXX}}

\begin{document}

\maketitle
\thispagestyle{firststyle}
\ifthenelse{\boolean{shortarticle}}{\ifthenelse{\boolean{singlecolumn}}{\abscontentformatted}{\abscontent}}{}

\dropcap{A}ctive matter exists over a wide range of spatial and temporal scales \citep{Lauga2009a,Ramaswamy2010,vicsekreview,RevModPhysMarchetti,RevModPhys.88.045006,Gompper2020}
from animal groups \citep{Cavagna2014,Berdahl2013} to robot swarms \citep{Rubenstein2014,Narayan2007,Deseigne2010}, 
to cell colonies and tissues \citep{Dombrowski2004,Zhang2010a,Lushi2014,Chen2017b,Yang2017tissue},
to cytoskeletal extracts \citep{Schaller2010,Sumino2012,Guillamat2016a,Wu2017,Guillamat2016a},
and man-made microswimmers \citep{Jiang2010a,Theurkauff2012,Bricard2013,Palacci2013,Yan2016}.
Constituent particles in active matter systems are driven out of thermal equilibrium at the individual level; they interact to develop a wealth of intriguing collective phenomena, including clustering \citep{Palacci2013,Zhang2010a,Theurkauff2012}, flocking \citep{Cavagna2010,Deseigne2010} , swarming \citep{Dombrowski2004,Zhang2010a}, spontaneous flow \citep{Lushi2014,Wu2017}, and giant density fluctuations \citep{Narayan2007,Deseigne2010}. Many of these observed phenomena have been successfully described by particle-based or continuum models\citep{Lauga2009a,Ramaswamy2010,vicsekreview,RevModPhysMarchetti,RevModPhys.88.045006,Gompper2020}, which highlight the important roles of both individual motility and inter-particle interactions in determining system dynamics.

Current active matter research focuses primarily on linearly swimming
particles which have a symmetric body and self-propel along one of the symmetry axes.
However, a perfect alignment between the propulsion direction and body axis is rarely found in reality.
Deviation from such a perfect alignment leads to a persistent curvature in the microswimmer trajectories; examples of such circle microswimmers include anisotropic artificial micromotors \citep{kummel2013,Du2020}, self-propelled nematic droplets \citep{Kruger2016,Lancia2019}, magnetotactic bacteria and Janus particles in rotating external fields \citep{Erglis2007,Han2017}, Janus particle in viscoelastic medium \citep{Narinder2018}, sperm and bacteria near interfaces \citep{Lauga2006,Riedel2005}. Chiral motility of circle microswimmers, as predicted by theoretical and numerical investigations, can lead to a range of interesting collective phenomena in circular microswimmers, including vortex structures \citep{Kaiser2013,Denk2016}, localization in traps \citep{Hoell2017}, enhanced flocking \citep{Liebchen2017}, and hyperuniform states \citep{Lei2019}. However, experimental verifications of these predictions are limited \citep{Riedel2005,Han2017}, a situation mainly due to the scarcity of suitable experimental systems.

Here we address this challenge by investigating marine algae $\it{Effrenium\ voratum}$
($\it{E.\ voratum}$) \citep{Jeong2014,LaJeunesse2018}. At air-liquid
interfaces, $E.\ voratum$ cells swim in circles via two eukaryotic flagella: a transverse flagellum encircling the cellular anteroposterior axis and a longitudinal one running posteriorly.
Over a wide range of densities, circling $E.\ voratum$ cells self-organize
into disordered hyperuniform states with suppressed density fluctuations at large lengthscales.
Hyperuniformity \citep{Torquato2003,Torquato2018} has been considered
as a new form of material order which leads to novel functionalities
\citep{Florescu2009,Xie2013,Man2013,Torquato2015,Ricouvier2019}; it has been
observed in many systems, including avian photoreceptor patterns \citep{Jiao2014}, amorphous ices \citep{Martelli2017}, amorphous silica \citep{Zheng2020},
ultracold atoms \citep{Lesanovsky2014}, soft matter systems \citep{Donev2005,Chremos2018,Tjhung2015,Weijs2015,Goldfriend2017,Weijs2017,Ricouvier2017,Wang2018}, and stochastic models \citep{Hexner2015,Hexner2017a,Hexner2017}.
Our work demonstrates the existence of hyperuniformity in active matter
and shows that hydrodynamic interactions can be used to construct hyperuniform
states.

\section*{Results}
$\it{E.\ voratum}$ belongs to family Symbiodiniaceae \citep{Jeong2014,LaJeunesse2018}.
Dinoflagellates in this family are among the most abundant eukaryotic
microbes found in coral reef ecosystems; they convert sunlight and
carbon dioxide into organic carbon and oxygen to fuel coral growth
and calcification \citep{Roth2014}. Cell motility of $\it{E.\ voratum}$
has been shown to be important for algal-invertebrate partnerships \citep{Aihara2019}, though
quantitative understanding of cell motility is still lacking.

\subsection*{Circular cell motion and associated flow field}

\begin{figure*}[tbhp]
\centering
\includegraphics[width=0.9\textwidth]{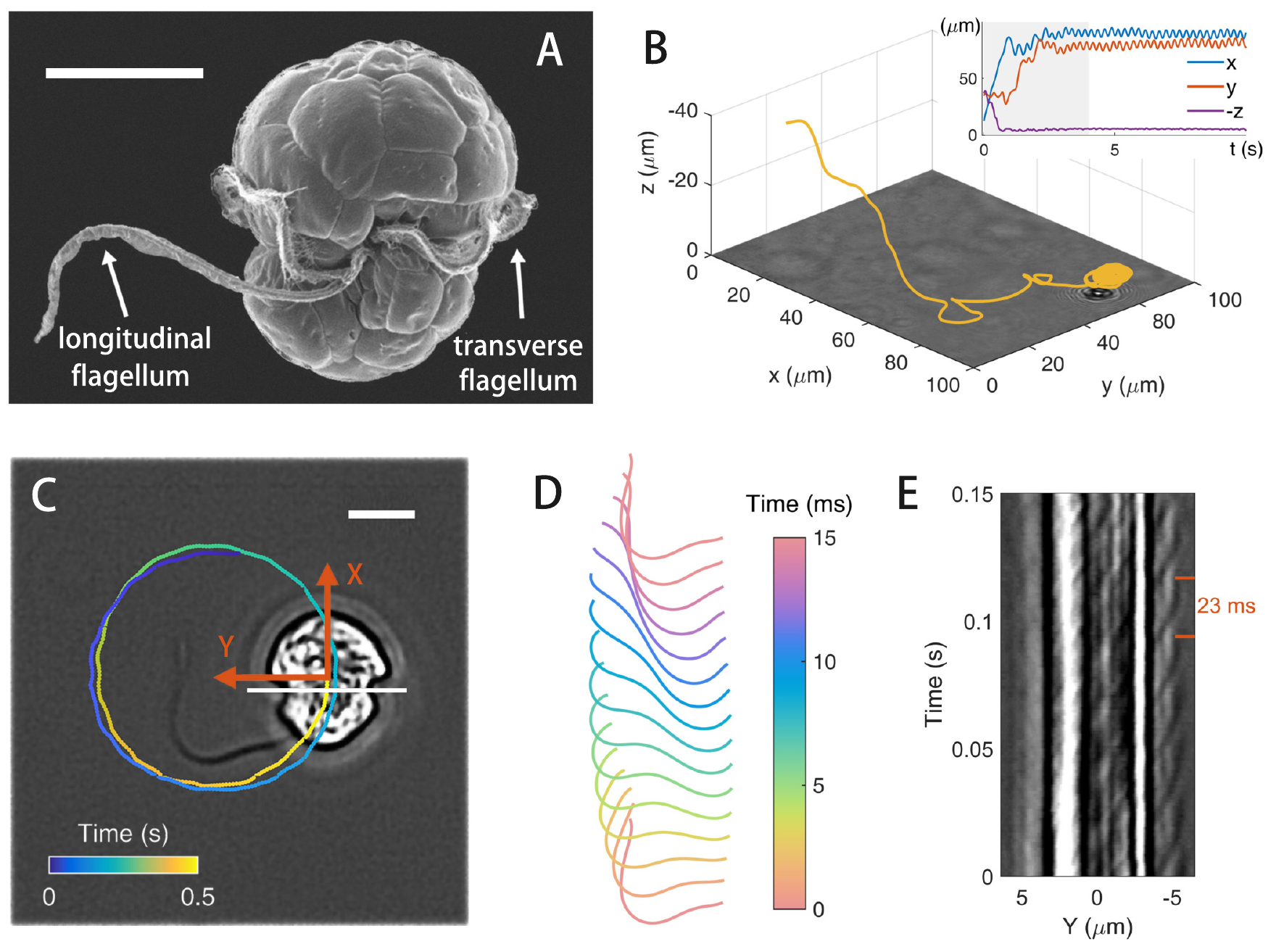}
\caption{
Cell motility and flagellar dynamics. (A) Scanning electron micrograph
of $\it{E.\ voratum}$ (courtesy of Sung Yeon Lee and Todd LaJeunesse \cite{Jeong2014}).
(B) 3D trajectory of a cell approaching an air-liquid interface (at $z=0$)
from the bulk. Time history of cell coordinates in the laboratory
frame ($x$, $y$, and $z$) is shown in the inset. (C) Circular
trajectory plotted on an optical image of a cell at the interface. Undulations in 
the trajectory reflect beating phases of the longitudinal flagellum.
A cell body frame (XY) is defined with X being the direction of
body axis. (D) Waveform of the longitudinal flagellum over a period.
(E) Kymograph to show transverse flagellum dynamics. Intensity profiles
of cell image are extracted along the white line (fixed in the cell body frame) in (C). Scale bars
in (A) and (C) are $5\ \mathrm{\mu m}$. See supporting videos S1
and S2 for cell motion and flagellar dynamics.
}
\label{fig:1}
\end{figure*}

We observe the cells at the air-liquid interface on an up-right microscope. As shown
in Fig. 1\textit{A} and \textit{C}, $\it{E.\ voratum}$ cells have an approximately elliptic
shape and are equipped with both longitudinal and transverse flagella
for motility. Away from interfaces, cells swim in helical trajectories which are typical
for motile algae, see Fig. 1\textit{B} and Movie S1. However, when cells get close to an air-liquid interface,
they adhere to the interface and start to move in circles; all cells move in
 the counter-clockwise direction when viewed from the air side of the air-liquid interface. In Fig. S3,
we show that cells also adhere to liquid-solid interfaces \citep{Ohmura2018} and estimate
the gap between cell and interface to be $0.3\ \text{\ensuremath{\mathrm{\mu m}}}$.
A typical counter-clockwise circular trajectory at air-liquid interface is plotted
on an optical image of a cell in Fig. 1\textit{C}, where we define the long
symmetric axis as the cell body direction, X coordinate. Typical cell circling radius, translation and angular velocities are
$\left\langle {a}\right\rangle = 11.6\ \text{\ensuremath{\mathrm{\mu m}}}$, $\left\langle {v_\mathrm{c}}\right\rangle = 180\ \text{\ensuremath{\mathrm{\mu m/s}}}$
and $\left\langle {\omega}\right\rangle =16.2\ \text{\ensuremath{\mathrm{rad/s}}}$, respectively. These motility
characteristics depend weakly on cell density and their variations
are quantified in Fig. S2\textit{B}.

As shown in Movie S2, the longitudinal flagellum produces a planar wave in a plane parallel to the air-liquid interface. Waveforms of the longitudinal flagellum in a period ($15\ \mathrm{ms}$) are shown in Fig. 1\textit{D}. The transverse flagellum sits in a groove \citep{Gaines2019,Fenchel2001,Miyasaka2004,Lee2008a}, as shown in Fig. 1\textit{A} and Movie S2; we cannot separate the flagellum's image from that of the cell body to extract all information about the flagellum's waveform. Instead, we extract intensity profiles from optical images along a line (fixed in the cell body frame) cutting through the transverse flagellum, cf. the white line in Fig. 1\textit{C}, and construct a kymograph from the extracted line profiles. As shown Fig. 1\textit{E}, the kymograph shows a wave propagation to the negative Y direction with a period about $23\ \mathrm{ms}$ and a wave speed of $124\ \mathrm{\mu m/s}$.

\begin{figure*}[h]
\centering
\includegraphics[width=0.8\textwidth]{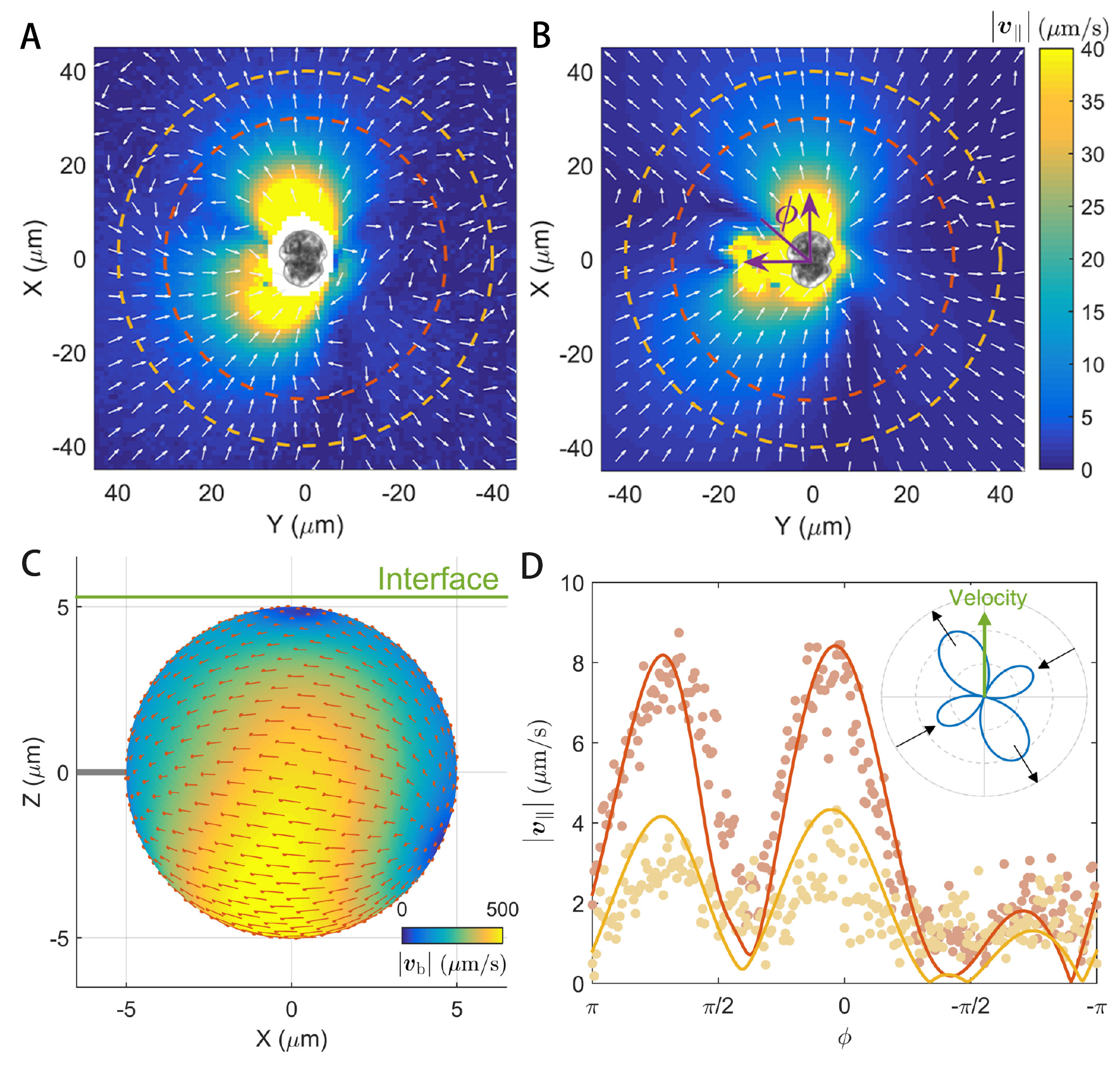}
\caption{
Mean in-plane flow field,  $\boldsymbol{v}_{\parallel}$, measured in experiment (A) and regularized Stokeslet model
(B). Flow speed ($\left|\boldsymbol{v}_{\parallel}\right|$) is represented by color and arrows show local flow direction.  Cell symmetry axis is oriented along the X axis and an angle from X direction is
defined as $\phi$ in (B). (C) An optimal slip flow pattern obtained from
our numerical procedure (see Text for details). The air-liquid interface is shown by a green line; the gap between
cell body and interface is set to be $0.3\ \mathrm{\mu m}$. 
(D) Angular dependence of the magnitude of the in-plane velocity at two radii, $30\ \mathrm{\mu m}$
and $40\ \mathrm{\mu m}$ dashed lines in (A) and (B), from experiments
(symbols) and numerics (lines). Angular dependence of far-field
flow speed (at the radius of $2000\ \mathrm{\mu m}$, computed from  regularized Stokeslet calculation)
is shown in the inset; the far-field flow is dominated by a pair of orthogonal 
pusher-puller dipoles (see SI for detail). Experimental data are measured from tracer motion around a cell with a  swimming speed  $v_{\mathrm{c}} = 201\ \mathrm{\mu m/s}$ and radius  $ a = 11.5\ \mathrm{\mu m}$, cf. Movie S3  .
}
\label{fig:2}
\end{figure*}

The swimming cell generates fluid flow (denoted by $\boldsymbol{v}$) in space. We measure the two flow components in  the plane of cell motion by tracking tracer particles, cf. Movie S3. The measured fields at different times are then averaged in the cell body (XY) frame. 
A typical averaged field, $\boldsymbol{v}_{\parallel}=v_{\text{X}}\hat{\text{X}}+v_{\text{Y}}\hat{\text{Y}}$, is plotted in Fig. 2\textit{A}. Though bearing some similarities with that of a  
source dipole, the field does not show any obvious (left-right or fore-after) symmetries, which are frequently found 
in the cases of straight swimmers \citep{Drescher2011}. To reproduce such
complex flow, we use a regularized Stokeslet model \citep{Ainley2008,Spagnolie2012,perkins1991hydrodynamic}. In the model,
the cell body is represented by a sphere with a radius of 5 $\mathrm{\mu m}$,
which is driven by both longitudinal and transverse flagella. As shown
in Fig. 1\textit{C-D}, the longitudinal flagellum has a conventional structure;
its planar waveform can be readily quantified and is faithfully represented
in the model. However, the transverse flagellum is hidden in the groove
and difficult to observe; its structure and driving mechanism are still
being debated \citep{Gaines2019,Fenchel2001,Miyasaka2004,Lee2008a}. We can measure the wave period and speed from Fig. 1\textit{E}, but the exact waveform of the transverse flagellum is unknown.
Due to this lack of information for the transverse flagellum, we represent
it in our model by a slip flow pattern on the cell surface, $\boldsymbol{v}_{\mathrm{b}}$. For a
given slip pattern, we use the regularized Stokeslet method to
compute cell (translation and rotation) velocities and a flow field
corresponding to the experimental result in Fig. 2\textit{A}, $\boldsymbol{v}_{\parallel}$. We vary the
slip flow pattern, $\boldsymbol{v}_{\mathrm{b}}$, and search for a pattern that optimizes the match
between numerical and experimental results of cell velocities and flow fields. A
resultant flow pattern, $\boldsymbol{v}_{b}$, from such a procedure is shown in Fig. 2\textit{C}
and maximal slip ($500\ \text{}\mu$m/s) occurs in the bright yellow region, which approximately corresponds to the location of the transverse
flagellum. With this slip pattern, we numerically generate the in-plane flow field
around a cell, shown in Fig. 2\textit{B}; two angular profiles of the in-plane flow speed ($\left|\boldsymbol{v}_{\parallel}\right|$) in Fig. 2\textit{D} show
good agreement between experiment and Stokeslet results. 
See SI for detailed discussions on the above procedure and results obtained in another cell, cf. Fig. S6.

\begin{figure*}[h]
\centering
\includegraphics[width=0.95\textwidth]{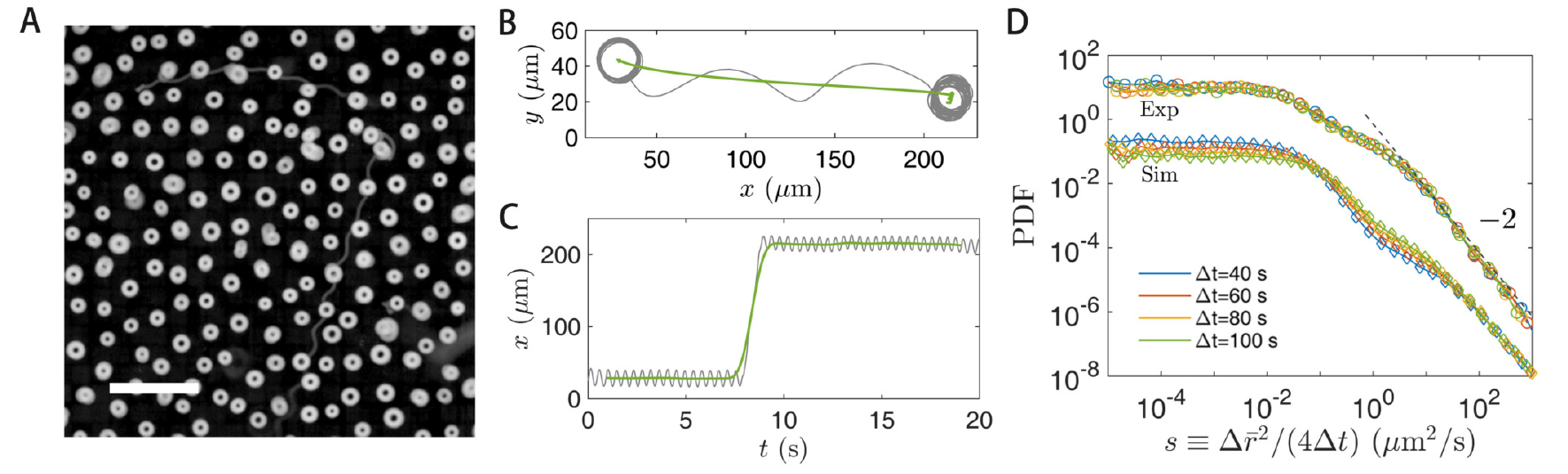}
\caption{
(A) Streak image of cell motion. Raw images are obtained at cell density
$178\ \mathrm{mm^{-2}}$ and averaged over 10 sec to produce the streak
image. Scale bar is $200\ \mathrm{\mu m}$. (B) Instantaneous cell coordinates (gray line, $\boldsymbol{r}\left(t\right)$) are averaged
over a sliding window of $2\ \mathrm{s}$ to highlight the non-circular motion (green line, $\boldsymbol{\bar{r}}\left(t\right)$). (C) Temporal history of instantaneous (gray) and window-averaged (green) coordinates. (D) Probability distribution functions of window-averaged cell
displacements (squared and normalized by time separation, $s\equiv\triangle\bar{r}^{2}/4\triangle t$ ) from experiment (circles) and model (squares) for four
time separations. The experiment and model results are in agreement; for presentation clarity, we rescale model results in (D) by a factor of 100.
}
\label{fig:3}
\end{figure*}

\subsection*{Non-circular cell motion}

Though circular motion is most frequently observed, cells
also exhibit rare non-circular motion. To demonstrate that, we show
a streak image of cell motion in Fig. 3\textit{A}. While majority of cells
move in circles and appear as white ''donuts'', the image also contains
rare long streaks, corresponding to rapid translational motion of
cells. The transition to non-circular motion is likely related to changes in the longitudinal flagellar dynamics, as depicted in Fig. S4 \citep{Ma2014,Wan2014}. 
We use a procedure with empirically chosen parameters to identify non-circular motion from instantaneous cell positions (see Fig. S5 and related discussions in SI for details). This procedure shows that durations of non-circular motion are usually less than a few seconds and that the occurring rate of non-circular motion is approximately $10^{-3}\ \rm{s^{-1}}\ \rm{cell^{-1}}$.

To quantify non-circular motion, we first average instantaneous cell coordinates, $\boldsymbol{r}\left(t\right)$,
over a sliding window of $2$ seconds ($\sim5.5$ circling period). As shown
in Fig. 3\textit{B-C}, circular motion is smoothed out in the window-averaged (green)
trajectory, denoted as $\boldsymbol{\bar{r}}\left(t\right)$. From
the window-averaged trajectories, we measure probability distributions of
cell displacements for different time separations $\triangle t$.
After squared displacements $\triangle\bar{r}^{2}\equiv\left(\bar{\boldsymbol{r}}\left(t+\triangle t\right)-\boldsymbol{\bar{r}}\left(t\right)\right)^{2}$
are normalized by the time separation $\triangle t$, all distributions
of $s\equiv\triangle\bar{r}^{2}/4\triangle t$ collapse onto a single
curve and exhibit a power-law scaling for large $s$, cf. Fig. 3\textit{D}. Similar probability distributions have been found under different density conditions, cf. Fig. S9. 

\begin{figure*}[h]
\centering
\includegraphics[width=0.95\textwidth]{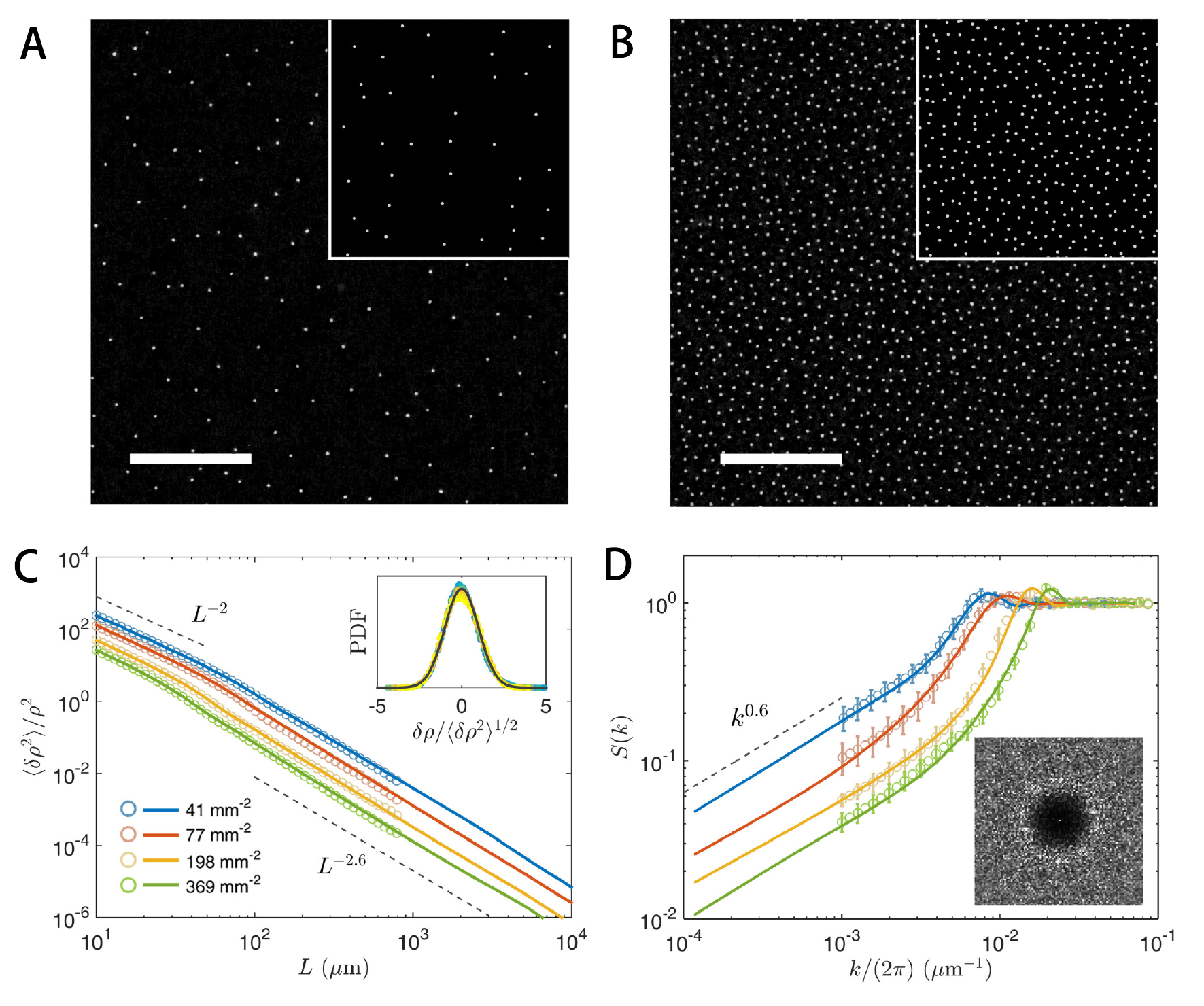}
\caption{
(A-B) Cell configurations at two densities ($41$
and $369\ \mathrm{mm^{-2}}$). Typical output of simulation is shown
in the inset. Scale bar is $500\ \mathrm{\mu m}$. (C) Cell density
fluctuations plotted against interrogation window size and (D) static
structure factors at four densities. Experimental and numerical results
are shown by light symbols and dark lines, respectively. The maximal measurement lengthscale is $800\ \mathrm{\mu m}$ in experiments and $10000\ \mathrm{\mu m}$ in simulations. The inset in (C) shows distributions of normalized density fluctuations measured at density $198\ \mathrm{mm^{-2}}$; color symbols are experimental data with different window sizes 
( $L=200$ to $800\ \mathrm{\mu m}$) and the thick line represent a normal distribution.
The inset in (D) shows a two-dimensional static structure factor measured 
at density $198\ \mathrm{mm^{-2}}$.
}
\label{fig:4}
\end{figure*}

\subsection*{Experimental observations of Hyperuniform states}
We next investigate collective states of interacting cells at the air-liquid interface. Cells in the bulk suspension swim to adhere at the interface in the first few minutes of experiments and this leads to a random initial distribution of cells at the interface, with a cell density $\rho$. Then, cells at the interface slowly self-organize into a steady 
state after a relaxation period about several thousand seconds, cf. Fig. S11. We measure static and dynamic properties of 
these steady states. Fig. 4\textit{A-B} show typical instantaneous configurations from two experiments, see also  Movie S4-5. Though no obvious order can
be detected in these configurations, spatial distribution of cells
appears to be quite uniform at large length scales. Quantitatively, from instantaneous cell positions  $\boldsymbol{r}^{\left(j\right)} \left(t\right)$, 
we compute density fluctuations for square interrogation windows of different sizes $L$. For a given window size, we find that density fluctuations follow a Gaussian distribution, as shown in the inset of Fig. 4\textit{C}. Variances of density fluctuations are plotted against the window size in Fig. 4\textit{C}; data follow the scaling determined
by the central limit theorem $\left\langle \delta\rho^{2} \right\rangle\sim L^{-2}$ at small scales and decay faster with the window size at large scales: $\left\langle \delta\rho^{2} \right\rangle\sim L^{-2.6}$. Similar physics is also
reflected by the static structure factors, $S\left(k\right)=\left\langle \frac{1}{N}\left|\sum_{j=1}^{N}\exp\left(-i\boldsymbol{k\cdot r}^{\left(j\right)}\right)\right|^{2}\right\rangle $, where $N$ is the total number of observed cells. In Fig. 4\textit{D}, $S\left(k\right)$
shows a liquid-like peak in large $k$ region and a scaling $S\left(k\right)\sim k^{0.6}$ for
small $k$. Lengthscale corresponding to liquid-like peaks matches
approximately to the transition length between $\left\langle \delta\rho^{2} \right\rangle\sim L^{-2}$
and $\left\langle \delta\rho^{2} \right\rangle\sim L^{-2.6}$ scalings. The same
scalings, $\left\langle \delta\rho^{2} \right\rangle\sim L^{-2.6}$ and $S\left(k\right)\sim k^{0.6}$,
are found for different cell densities. 
Increasing the cell density leads to a decrease of $S\left(k\right)$ for small $k$ and shifts the
liquid-like peak to larger $k$, as shown by Fig. 4\textit{C-D}. 
To check the robustness of observed hyperuniformity, we also use window-averaged cell position, $\boldsymbol{\bar{r}}\left(t\right)$ defined in Fig. 3\textit{B}, to compute density fluctuations and obtain similar results, shown in Fig. S11. 

\begin{figure*}[h]
\centering
\includegraphics[width=0.95\textwidth]{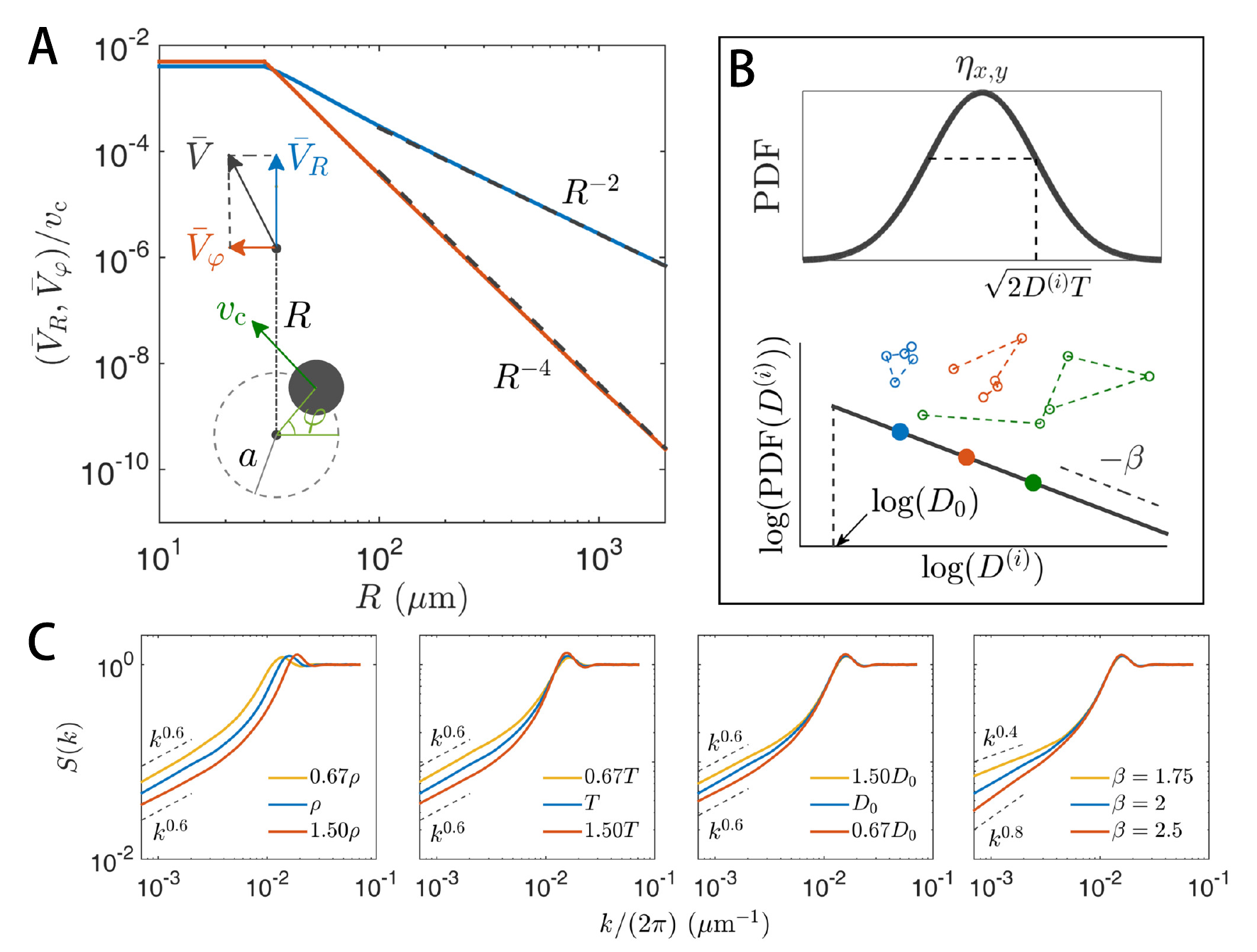}
\caption{
(A) Radial and tangential components of period-averaged flow field
around a swimming cell. Flow components are defined in the inset, showing
a cell circling with a radius $a$ and a velocity $v_{{\rm c}}$. Flow strength is capped at $R=30\mathrm{\mu m}$, which approximately corresponds to the minimal distance between circular centers of two cells in experiments. Black dash lines show the asymptotic behavior in the far-field(see text). (B) Normal distribution of jumping displacements $\boldsymbol{\eta}\left(D^{\left(i\right)}\right)$ (cf. Eq.[1]) and Pareto distribution of particle diffusivity ($D^{\left(i\right)}$) (cf. Eq.[2]). Three stochastic trajectories with different diffusivity are shown in the inset. (C) Effects of four model parameters on structure factors.
A default set of parameters is used unless specified: $\rho=198\ \mathrm{mm^{-2}}$,
$T=400\ \mathrm{s}$, $D_{0}=0.95\ \mathrm{\mathrm{\mu m^{2}/s}}$
and $\beta=2$. {}{}
}
\label{fig:5}
\end{figure*}

Beyond static structures, we also investigate the system's dynamic properties. To explore the possibility of local synchronization, spatial correlation functions of circling phase and velocity are computed; results in Fig. S10\textit{C-D} show that instantaneous cell motions are not spatially correlated, suggesting a weak interaction between cells. This is confirmed by Fig. 2 which shows that the flow velocity at the nearest cells ( at a distance $\sim 30-100 \  \mu\mathrm{m} $) is much smaller than cell swimming speed. Therefore, the cell-cell interaction is not strong enough to significantly affect the instantaneous cell motion. However, as shown below, this weak hydrodynamic interaction can modulate cell positions over a long time and its long-range nature leads to the formation of hyperuniform states.

\subsection*{Particle-based model for hyperuniformity} We construct a numerical
model to illustrate the origin of observed hyperuniformity. Dynamics in our experiments evolves over a time scale that is much longer than cell circling periods  ($\sim0.4\ \mathrm{s}$). This separation of time scales allows us to build a temporally coarse-grained (over a few circling periods)
model to capture the emergence of hyperuniformity without fully resolving
fast circular cell motion \citep{Nagai2015}. Therefore, particle coordinates in our
model represent window-averaged cell positions in experiments ($\boldsymbol{\bar{r}}\left(t\right)$ in Fig. 3\textit{B}) and particles interact through period-averaged
flow field. Beyond flow advection, our model also includes stochastic non-circular particle motion and uses the following equation to determine the displacement of the $i$th
particle at time $n\tau$ during a time step $\tau$:
\begin{equation}
\begin{array}{l}
\boldsymbol{\bar{r}}^{\left(i\right)}\left(\left(n+1\right)\tau\right)-\boldsymbol{\bar{r}}^{\left(i\right)}\left(n\tau\right)=\\
\underset{j\neq i}{\sum}\boldsymbol{\bar{V}}\left(\boldsymbol{\bar{r}}^{\left(i\right)}\left(n\tau\right)-\boldsymbol{\bar{r}}^{\left(j\right)}\left(n\tau\right);v_{\mathrm{c}}^{\left(j\right)}\right)\tau\\
+\boldsymbol{\eta}\left(D^{\left(i\right)}\right)\delta\left(\text{mod}\left(n,p \right),s^{\left(i\right)}\right)
\end{array}.\label{eq:3}
\end{equation}

The period-averaged flow field in Eq. [\ref{eq:3}], $\boldsymbol{\bar{V}}\left(\boldsymbol{R};v_{\mathrm{c}}\right)$, is calculated by the regularized Stokeslet method for a cell circling with a radius $a = 10\  \mathrm{\mu m}$ and velocity $v_{\mathrm{c}}$. As shown in Fig. 5\textit{A}, $\boldsymbol{\bar{V}}\left(\boldsymbol{R};v_{\mathrm{c}}\right)$ has an outgoing component in the far field; see Materials and Methods for detailed discussions on the period-averaged flow field. In our model, parameter $v_{\mathrm{c}}^{\left(i\right)}$ is sampled from an experimentally determined distribution of cell velocity in Fig. S2.

The second term on the right-hand-side of Eq.[\ref{eq:3}] represents stochastic jumps. The Kronecker delta function $\delta$$\left(\right)$ dictates that adjacent random jumps for a given particle are temporally separated by $p$ (an integer constant) time steps, defining a waiting time $T=p\tau$. Specifically, the $i$th particle jumps at time $n\tau$ if $\text{mod}\left(n,p\right)=s^{\left(i\right)}$; mod$\left(\right)$ represents modulo operation and $s^{\left(i\right)}$ is an integer constant between $0$ and $p-1$, randomly assigned to all particles. Components of jumping displacements $\boldsymbol{\eta}\left(D^{\left(i\right)}\right)$ are independently drawn from a normal distribution with a standard deviation $\sqrt{2D^{(i)}T}$, as shown in Fig. 5\textit{B}. Parameter $D^{\left(i\right)}$ is the diffusivity for the $i$th particle and drawn from a Pareto distribution with a cut-off value $D_{0}$ and a power index $\beta$ :
\begin{equation}
f\left(D;D_{0},\beta\right)=\begin{cases}
\left(\beta-1\right)\frac{D_{0}^{\beta-1}}{D^{\beta}} & D\geq D_{0}\\
0 & D<D_{0}
\end{cases},\label{eq:fd}
\end{equation}
where $\beta=2$ unless stated otherwise. 

In a 2D periodic domain of size $L_{\mathrm{max}}=20\ \mathrm{mm}$, we simulate $\rho L_{\mathrm{max}}^{2}$ particles following Eq. [\ref{eq:3}]. For a given experimental condition, the cut-off diffusivity $D_{0}$ and the waiting
time $T$ are varied to match simulation results to experiments. The obtained values for $D_{0}$ ($\sim 1\ \mathrm{\mathrm{\mu m^{2}/s}}$) and $T$ ($\sim 500\ \mathrm{s}$) are listed for different cell densities in Table SIII. 

We measure density fluctuations after randomly initialized particles evolve to a steady state. As shown in Fig. 4\textit{C-D}, our simulations
can generate hyperuniform states and quantitatively reproduce measured
density fluctuation $\delta\rho^{2}$ and structure
factor $S\left(k\right)$, highlighting scaling laws $\delta\rho^{2}\sim L^{-2.6}$
and $S\left(k\right)\sim k^{0.6}$ for hyperuniformity. Distribution
functions of cell displacements are also well reproduced in Fig. 3\textit{D}.
The Pareto distribution of $D^{\left(i\right)}$ with a power index
$\beta=2$ leads to the observed power-law distribution for large
displacements. 

We systematically study effects of model parameters; results are
shown in Fig. 5\textit{C}. An increase of particle density $\rho$ shifts
the liquid-like peak to higher $k$ and leads to a decrease of density
fluctuation for small $k$, which mirrors experimental results in
Fig. 4\textit{D}. Deceasing the waiting time $T$ introduces more fluctuations into
the system and lead to an increase in $S\left(k\right)$ for small
$k$, cf. the second panel in Fig. 5\textit{C}. Smaller cut-off diffusivity $D_{0}$ allows fewer particles with large $D^{\left(i\right)}$; this  leads to fewer stochastic jumps with large displacement, less fluctuations in large scales and smaller $S\left(k\right)$ values. As shown in the fourth panel in Fig. 5\textit{C}, the power index
of the Pareto distribution, $\beta$ in Eq. [\ref{eq:fd}], can change
power-law scaling of $S\left(k\right)$ at small $k$.
A larger $\beta$ means fewer particles with large diffusivity; 
this leads the system to approach the strong hyperuniformity limit \citep{Torquato2018} and a larger exponent in $S\left(k\right)$ is observed. Fig. 5\textit{C} also demonstrates that variations in $\rho$,
$T$ and $D_{0}$ have relatively weak effects on the scaling exponent for
small $k$ in our observation window \citep{Kim2018a}.

\section*{Discussion}
In summary, we have studied individual motility
and collective dynamics in marine algae \textit{E.\ voratum}.
Cells swim in circles at the air-liquid interface with a longitudinal flagellum and a transverse one. Combining experimental measurements and the regularized
Stokeslet method, we showed that period-averaged flow generated by cells
has a long-ranged and out-going radial component that disperses cells uniformly 
and leads to a disordered hyperuniform state. Stochastic cell motion with a power-law displacement distribution  \citep{Kim2018a} also plays 
an important role in determining the properties of density fluctuations.  

Regularized Stokeslet results in Fig. 2 can be used to clarify the current confusion on the contributions of two flagella to propulsion in \textit{E.\ voratum} \citep{Gaines2019,Fenchel2001,Miyasaka2004}. For that, 
we measure the stalled force and torque with both flagella or only one of them functioning; as shown in SI, 
the longitudinal flagellum provides about 30\% of the total torque and less than 10\% of the force. The regularized Stokeslet
method is also used to calculate 3D period-average flow and show that self-generated flow can lead to directed nutrient/particle transport around cells, cf. Fig. S8. These results provide novel insight into
the ecological function and evolutionary traits of two flagella in this ecologically important dinoflagellate \citep{Jeong2014,Roth2014,LaJeunesse2018,Aihara2019}. 

Previous studies have shown that hydrodynamic interactions lead to
interesting self-organization \citep{Goldfriend2017}. For example, bacteria, when
oriented perpendicularly to an interface, can generate inward flow that
assembles bacterial cells into compact crystals at the interface \citep{Petroff,Chen2015}. In contrast,
circling $\it{E.\ voratum}$ cells in current work generate repulsive interactions. This can 
be understood from the far-field instantaneous flow in the plane cell reside. As shown in the inset of Fig. 2\textit{D} and analysis in SI,
out-going flow in the far-field is stronger than its in-coming counterpart. This leads to a period-average repulsive interaction between
cells in Fig. 5\textit{A}. The same hydrodynamic mechanism may underlie 
the formation of sperm vortex in \cite{Riedel2005}. Our work suggests 
a novel mechanism of using average along circular trajectories to generate isotropic hydrodynamic interactions between force-free microswimmers. Such isotropic hydrodynamic interactions in chiral active matter are different from anisotropic dipolar interactions in conventional systems with linearly swimming particles and may produce new collective phenomena.

Fig. 4 shows that hyperuniformity is observed under different cell concentration conditions with a similar scaling exponent. Such a density independence has also been observed in other hyperuniform systems with long-range interactions. In a one-component plasma, particles with the same electrostatic charge interact with repulsive Coulomb potential which imposes energy penalty on density fluctuations and leads to hyperuniformity under all particle density conditions \citep{Torquato2018}. In sedimentation system of irregular objects \cite{Goldfriend2017}, falling objects interact via long-range hydrodynamic (force monopole) flow and objects’ irregular shape leads to an anisotropic response to the local flow. A combination of the long-range interaction and anisotropic response in this system produces hyperuniformity with a density-independent exponent.

Hyperuniformity has been observed in many systems exhibiting absorbing-state transition 
\citep{Weijs2015,Hexner2015,Hexner2017,Hexner2017a,Lei2019,Lei2019a}. For example, 
a recent numerical work simulated a system of active particles which self-propel 
in circles (alike $\it{E.\ voratum}$ cells here) but interact via short-range repulsive forces; 
hyperuniformity in this system was only observed in high-density active states \cite{Lei2019}. 
In contrast, $\it{E.\ voratum}$ cells self-organize into hyperuniform states under all densities, thanks to the long-range 
nature of hydrodynamic interactions \citep{Torquato2015,Torquato2018}. Phoretic interactions
in synthetic active matter system are also known to be long-ranged; both attractive and repulsive phoretic interactions 
have been realized \citep{Palacci2013,Moran2017}. These long-range interactions may provide a promising avenue 
to generate novel hyperuniform materials with active matter.

\matmethods{\subsection*{Cell growth and imaging procedure}
Spices $\it{E.\ voratum}$ cells are cultured in
artificial seawater with F/2 medium in a 100 ml flask which is placed in an incubator (INFORS HT Multitron pro, Switzerland) 
 at $20^{\circ}{\rm C}$. We use a daily light cycle which consists of 12 hours of cool light with an intensity of 2,000
Lux and 12 hours in dark. The algal cells
in the experiments are in an exponential phase after 14-day growth
and observed a few hours after the light period starts, when cells
showed excellent motility (see Fig. S1, \citep{Jeong2014,LaJeunesse2018}).
During experiments, cell culture is placed in a disk-shape chamber
fabricated by cover glass and plastic gasket (8-mm in diameter). Cells gather and form a mono-layer
at the air-liquid interface. Cell motion in the central region of the sample is recorded
by a high-speed camera (Basler acA2040-180km, 4 M pixel resolution)
mounted on an up-right microscope (Nikon Ni-U) with a 4\texttimes{} or 40\texttimes{} 
magnification objective; the acquisition rate varies from 50 to 850
frame/s. To measure fluid flow, milk (Deluxe Milk, Mengniu) is added
to provide passive flow tracers (diameter 1$\sim$2 $\mathrm{\mu m}$).
Holographic imaging technique is used to measure 3D cell motion. 

\subsection*{Particle-based model}
To obtain period-averaged flow field in Fig. 5\textit{A}, we
use the regularized Stokeslet results from Fig. 2 to compute instantaneous
flow fields around a cell and average computed fields in the lab frame over cell
positions in a circling period. The flow field is capped at $R=30\ \mathrm{\mu m}$, which approximately corresponds to the minimal distance between circular center of two cells in experiments. In simulation, we interpolate data in Fig. 5\textit{A} to find two flow components at any separation; asymptotic expressions for flow components (Eq. [S21]) with a cap at at $R=30\ \mathrm{\mu m}$ can also reproduce experimental data, cf. Fig. S12 for details. To reduce the computing load, interacting flow is assumed to be zero beyond a cutoff length of 
 $20000\ {\rm \mu m}$, which is twice the maximal length of computed density fluctuations $\langle \delta\rho^{2} \rangle$ and structure factor $S\left(k\right)$ in simulations.

In our model, time step $\tau$ is set to be 10 sec, during which typical particle 
displacements ($\sim 1\ \mathrm{\mathrm{\mu m}}$) are much smaller than typical 
particle separations ($\sim 50\ \mathrm{\mathrm{\mu m}}$). Analysis of experiments show 
that durations of random non-circular motion are less than 10 sec, cf. Fig. S5\textit{D}; such events 
occur within a single time-step in simulation.

Particle positions in simulations represent temporally 
averaged cell positions $\boldsymbol{\bar{r}^{(i)}}\left(t\right)$; we add a random circling phase (cf. Fig. S10\textit{C}) to 
each particle position to obtain instantaneous cell positions, $\boldsymbol{{r}^{(i)}}\left(t\right)$, 
which are used to compute density fluctuations.

}

\showmatmethods{} 

\acknow{We acknowledge financial supports from National Natural Science Foundation of China Grants (12074243, 11774222 and 32071609) and from the Program for Professor of Special Appointment at Shanghai Institutions of Higher Learning (Grant GZ2016004). We thank the Student Innovation Center at Shanghai Jiao Tong University for support. }

\showacknow{} 

\bibliography{select}

\end{document}